\documentclass[RNAAS]{aastex63}

\graphicspath{{./}{figures/}}

\begin{document}

\title{An Arecibo 327 MHz Search for Radio Pulsars and Bursts in the Dwarf Irregular Galaxies Leo A and T}

\correspondingauthor{Fronefield Crawford}
\email{fcrawfor@fandm.edu}

\author[0000-0002-2578-0360]{Fronefield Crawford}
\affiliation{Department of Physics and Astronomy, Franklin and Marshall College, P.O. Box 3003, Lancaster, PA 17604, USA}

\author[0000-0002-8758-8139]{Kohei Hayashi}
\affiliation{Astronomical Institute, Tohoku University, Sendai, Miyagi, 980-8578, Japan}
\affiliation{National Institute of Technology, Ichinoseki College, Takanashi, Hagisho, Ichinoseki, Iwate, 021-8511, Japan}

\begin{abstract}
We have conducted an Arecibo 327 MHz search of two dwarf irregular galaxies in the Local Group, Leo A and T, for radio pulsars and single pulses from fast radio bursts and other giant pulse emitters. We detected no astrophysical signals in this search, and we estimate flux density limits on both periodic and burst emission. Our derived luminosity limits indicate that only the most luminous radio pulsars known in our Galaxy and in the Magellanic Clouds (MCs) would have been detectable in our search if they were at the distances of Leo A and T. Given the much smaller stellar mass content and star formation rates of Leo A and T compared to the Milky Way and the MCs, there are likely to be few (if any) extremely luminous pulsars in these galaxies. It is therefore not surprising that we detected no pulsars in our search.
\end{abstract}


\section{Introduction} 

The detection and study of extragalactic pulsars offers the rare opportunity to probe the high end of the pulsar luminosity function in another galaxy. Pulsar searches are also sensitive to bursting sources, such as fast radio bursts (FRBs; \citealt{lbm+07}), and pulsed radio signals originating from a nearby galaxy can be used to probe the ionized intergalactic medium and the interstellar medium of the host galaxy. A number of searches for pulsars in nearby galaxies have been attempted (e.g., \citealt{mc03, rsh+13, mv16, vmk+20}). However, to date, the only confirmed extragalactic radio pulsar discoveries have been in the Magellanic Clouds (MCs), the closest galaxies to our own \citep{mfl+06, rcl+13}. Leo A and T are in the Local Group and are classified as dwarf irregular galaxies~(dIrrs), which are gas-rich dwarfs with active star formation. Since many dIrrs are isolated (i.e., outside of the realm of the Milky Way and M31), their star formation activities are thought not to be significantly interrupted by external effects such as tidal disruption. The star formation rates (SFRs) of dIrrs are much higher than those of dwarf spheroidal galaxies, which they resemble but which have little gas and no current star forming activity. However, the SFRs of Leo A and T \citep{rbo+08, kk13} are still several orders of magnitude smaller than the SFRs of the Milky Way and MCs \citep{hz09} owing to their much smaller masses. We selected Leo A and T to search from the list of dwarf galaxies in or around the Local Group that were cataloged by \citet{m12}. These two galaxies were selected owing to their sky positions (visible from Arecibo), their relatively small distances from Earth, and their small angular sizes (each galaxy fits within a single 327 MHz Arecibo beam). Leo A is the larger of the two galaxies and so should harbor a greater number of pulsars. However, Leo T is at half the distance to Leo A (Table \ref{tbl-1}) and hence weaker pulsars would be more easily detectable there. Neither Leo A or T appears to have been previously searched for pulsars. 

\section{Observations and Analysis}

We used the Arecibo 305-m telescope to search both Leo A and T for periodic emission from pulsars and for single pulses from FRBs and other giant pulse emitters. We used a center frequency of 327 MHz and the PUPPI backend, with an effective bandwidth of 68.75 MHz split into 2816 channels. Each channel was 8-bit sampled every 81.92 $\mu$s, with total integration times of 40 min (for Leo A) and 31 min (for Leo T). The Arecibo beam size at 327 MHz is 15~arcmin, much larger than the angular sizes of the two galaxies (see Table \ref{tbl-1}). Thus, in each case the entire galaxy could be observed with a single telescope pointing. The data were processed using the PRESTO package \citep{r01}\footnote{\url{http://www.cv.nrao.edu/~sransom/presto/}} for the periodicity search. The data were inspected for radio frequency interference and then dedispersed at trial dispersion measures~(DMs) between 0 and 3000 pc cm$^{-3}$. Each dedispersed time series was Fourier transformed, and the resulting power spectrum was filtered to remove red noise before it was harmonically summed and searched for significant periodic signals. No acceleration search was performed, which limited our sensitivity to any highly accelerated binary pulsars. For the single pulse search, we dedispersed the data at trial DMs between 0 and 1250 pc cm$^{-3}$ and applied the HEIMDALL \citep{b12}\footnote{\url{https://sourceforge.net/projects/heimdall-astro/}} package to identify significant pulses in each dedispersed time series. The detected pulses were then passed to the single-pulse classifier FETCH \citep{aab+20}\footnote{\url{https://github.com/devanshkv/fetch/}}, and each detected pulse was assigned a probability by FETCH of being a real, astrophysical signal. 

\section{Results and Discussion} 

No periodic or single-pulse signals of astrophysical origin were detected above a signal-to-noise ratio of 7. Our 327 MHz flux density limit to single pulses was $\sim 0.30$~Jy for a 1 ms pulse. Our 327 MHz flux density limits for periodic emission from radio pulsars were $\sim 0.047$ and $\sim 0.053$ mJy for Leo A and T, respectively (Table \ref{tbl-1}), assuming a 5\% pulsed duty cycle. This was obtained from the radiometer equation and was checked empirically using several test pulsar observations. Translating these periodicity flux density limits to radio luminosity limits using the distance to each galaxy (see Table \ref{tbl-1}) gives 327 MHz luminosity limits of 30000 and 9200~mJy kpc$^{2}$ for Leo A and T, respectively. These luminosity limits are higher than the estimated 327 MHz luminosities for the vast majority of pulsars listed in the ATNF catalog \citep{mht+05}\footnote{\url{https://www.atnf.csiro.au/research/pulsar/psrcat/}}. The 327 MHz luminosities for catalog pulsars were estimated using their measured 1400 MHz flux densities, their cataloged distances, and an assumed spectral index of $-1.4$ in each case (the expected average for the underlying pulsar population; see \citealt{blv13}). A total of 6 cataloged Galactic pulsars have estimated 327 MHz luminosities greater than our Leo T limit of 9200~mJy kpc$^{2}$, and one of these lies above our Leo A limit of 30000 mJy kpc$^{2}$. For cataloged pulsars in the MCs, where the distances are better known, we find that one pulsar in the SMC has an estimated 327 MHz luminosity greater than 9200~mJy kpc$^{2}$, and no pulsars have a luminosity greater than 30000~mJy kpc$^{2}$. Despite the uncertainties in these estimates, it is clear that the luminosity limits for our periodicity search are near the limit of detectability for even the most luminous pulsars currently known. The much smaller stellar mass content and SFRs of both Leo A and T compared to the Milky Way and the MCs suggest that few (if any) extremely luminous pulsars are likely to be present in Leo A and T. It is therefore not surprising that no such pulsars were detected in our search.

\begin{deluxetable}{lcc}
\tablecaption{Observed Galaxies and and Periodicity Search Limits\label{tbl-1}}
\tablehead{
\colhead{Name} \vspace{-0.0cm} & \colhead{Leo A} & \colhead{Leo T}
}
\startdata
Right Ascension (J2000) (hh:mm:ss)  & 09:59:26  &  09:34:53         \\
Declination (J2000) (dd:mm:ss)      & +30:44:47 & +17:03:05         \\
Distance (kpc)                      & 798       & 417               \\
Angular Size (arcmin)               & 5.1 x 3.1 & 1.4 x 1.4         \\
Integration Time (s)                & 2395      & 1885              \\
327 MHz Flux Density Limit (mJy)    & 0.047     & 0.053             \\
327 MHz Luminosity Limit (mJy kpc$^{2}$) & 30000 & 9200 
\enddata
\tablecomments{From \citet{m12}.} 
\end{deluxetable}

\bibliography{ch21}{}
\bibliographystyle{aasjournal}
\end{document}